\documentclass[copyright]{eptcs}

\usepackage[latin1]{inputenc}
\usepackage{ae}
\usepackage{amssymb}

\usepackage{tikz}
\usetikzlibrary{petri}\tikzset{
  >=latex,
  every place/.style={minimum size=6mm},
  every transition/.style={minimum size=6mm},
  token distance=5pt,
}

\tikzstyle{cover}=[fill=white,opacity=.6,text opacity=1,inner
  sep=.1em,outer sep=.2333em]

\def\defeq{\mathrel{\smash{\stackrel{\mbox{{\tiny\rm df}}}{=}}}}
\def\round{\mathop{\mathsf{round}}}
\def\max{\mathop{\mathsf{max}}}
\def\min{\mathop{\mathsf{min}}}
\def\GBF #1{\ensuremath{\mathsf{#1}}}
\def\ie{{\emph{i.e.}}}
\def\eg{{\emph{e.g.}}}
\def\diff{\mathbin{\triangleright}}

\providecommand{\event}{G.Ciobanu, M.Koutny (Eds.): Membrane Computing and\\
  Biologically Inspired Process Calculi (MeCBIC 2010)\\
  EPTCS Proceedings, 2010.}

\title{Qualitative modelling and analysis of\\
       regulations in multi-cellular systems using\\
       Petri nets and topological collections}
\author{Jean-Louis Giavitto
 \qquad\qquad Hanna Klaudel
 \qquad\qquad Franck Pommereau
\institute{IBISC, University of Évry}
\institute{Tour Évry 2, 523 place des terrasses de l'Agora,
  91000 Évry, France}
\email{\{giavitto,klaudel,pommereau\}@ibisc.univ-evry.fr}}

\def\titlerunning{Qualitative modelling and analysis of
                  multi-cellular systems}
\def\authorrunning{J.-L.~Giavitto, H.~Klaudel \& F.~Pommereau}

\begin{document}
\maketitle

\begin{abstract}
      In this paper, we aim at modelling and analyzing the regulation
      processes in multi-cellular biological systems, in particular
      tissues.
      The modelling framework is based on interconnected logical
      regulatory networks (\emph{à la} Ren\'e Thomas) equipped with
      information about their spatial relationships. The semantics of
      such models is expressed through colored Petri nets to
      implement regulation rules, combined with topological
      collections to implement the spatial information. 
      Some constraints are put on the the representation of spatial
      information in order to preserve the possibility of an
      enumerative and exhaustive state space exploration.
      This paper presents the modelling framework, its semantics, as
      well as a prototype implementation that allowed preliminary
      experimentation on some applications.
\end{abstract}

\section{Introduction}

Regulation processes are the corner stone to understand many aspects
of biological systems. Regulations occur at many levels: transcription
and translation of the genetic material, protein modifications, etc.
They define complex networks of interactions that cannot be easily
understood without resorting to formal modelling and automated
analysis. The generalized logical formalism initially proposed by
Ren\'e Thomas in the 70s~\cite{thomas73a,thomas91a,thomas95}, is a
discrete modelling formalism that has proved to be an effective way to
capture regulation processes and analyze them. It has been
successfully applied to the study of a variety of regulatory networks
comprising relatively large numbers of
components~\cite{saez07,sanchez08}. This formalism however does not
provide any modelling device to specifically address the question of
multi-cellular systems, where the regulatory networks of cells can
influence each other in a way that is dependent on the spatial
relationships between the cells.

This paper is a step toward providing a modelling framework for the
regulation in multi-cellular systems, in particular tissues, taking
into account cells migration, division and apoptosis (death).  This
framework will be applied to the analysis of systems such as
developmental processes, invasive cancers, plant growth, etc.

In such a modelling framework, we would like to preserve the ability
to perform model-checking based analysis in order to be able to assess
causality-related properties and observe rare events, which usually
cannot be obtained through simulation. This constrains the possible
choices for modelling spatial information. In particular, their should
be a finite number of possible spatial evolutions from a given
configuration and they should be enumerable. Moreover, each spatial
configuration should be represented in a normalized form, allowing the
recognition of two identical configurations. For instance, floating-point
positions are not a possible solution, instead, we shall use solutions
based on discrete representations. Finally, we would like to be able
to obtain easily a graphical representation of a cell population, \ie,
the chosen representation of spatial information should be compatible
with existing graphical rendering techniques.

\paragraph{Contributions.}

Our starting point is a modelling approach developed
in~\cite{chaouiya06a,chaouiya09a} and extended later to introduce
modularity~\cite{ckp10,jb09}. This approach itself is based on logical
regulatory networks~\cite{thomas73a,thomas91a,thomas95} that allow for
modelling regulation within a biological system. A Petri net semantics
is provided in order to analyze the various properties of so modelled
systems. In~\cite{ckp10,jb09}, modularity allows an easier modelling
of multi-cellular systems, each cell being represented by a module.
However, the spatial relationship between the modules within a model
is represented in a very abstract way, with no link to any kind of
geometrical or topological information.

Our main contribution in this paper is twofold. First, we propose a
way to specify such information in a very flexible way, without
significantly increasing the complexity of the original framework.
Second, we define devices to model efficiently the spatial
transformations related to the apoptosis, migration and division
processes. The former goal is achieved by decoupling the regulation
rules in the model from the spatial information and transformations.
Both aspects being linked through a standardized interface, allowing
the modeler for using various approaches to spatial representation.
The latter goal is achieved by a careful design of this interface in
order to enable a smooth bi-directional communication between the two
parts of a model.

Then, we present two approaches to the modelling of spatial
information. One is based on predefined grids and another is based on
bounded-degree graphs. Both methods have pros and cons, depending on
the application domain of interest. Finally, our framework is given a
Petri net semantics that is implemented in a prototype, allowing to
prove the feasibility of the approach and to run preliminary
experiments on simple applications.

Notice that the approach presented in this paper is applicable only if
one sort of module is considered at the same time, \ie, when all the
cells in a tissue are of the same kind. The extension to take into
account multi-sorted systems is quite straightforward, but the
resulting notations are much more complex. Our prototype implementation
actually does not have any such limitation. However, for this paper,
an intuitive and simpler presentation has been preferred.

\paragraph{Outlines.}

The next section introduces the background of our work, in particular
the modelling framework we start from. Our contributions are then
presented in sections~\ref{sec:spatial} and~\ref{sec:contrib}, the
former being dedicated to the modelling of spatial information while
the latter presents the extended modelling framework. The paper ends
with a conclusion and a discussion about future works.

\section{Background}

\subsection{Logical regulatory networks}

\begin{figure}
\begin{center}
\begin{tikzpicture}
\node[place] (A) at (0,1) {$A$};
\node[place] (B) at (1,1) {$B$};
\node[place] (C) at (0,0) {$C$};
\draw[->](A)--(B);
\draw[->](B)--(C);
\draw[-|,shorten >=1pt](C)--(A);
\node[right] at (2.5,1.1) {$K_A: x_C \to \{0,1\} \defeq 1 - x_C$};
\node[right] at (2.5,.5) {$K_B: x_A \to \{0,1\} \defeq x_A$};
\node[right] at (2.5,-.1) {$K_C: x_B \to \{0,1\} \defeq x_B$};
\end{tikzpicture}
\end{center}
\caption{A simple regulatory network with three components, $A$, $B$
  and $C$, such that $A$ is an activator for $B$, $B$ is an activator
  for $C$, and $C$ is an inhibitor for $A$.}
\label{fig:simple-regnet}
\end{figure}
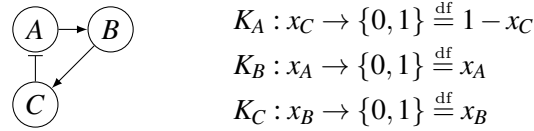

A logical \emph{regulatory network} is usually depicted as a graph
whose nodes are \emph{regulatory components}, for instance genes or
proteins, and whose arcs indicate how each component is influenced by
others. A simple regulatory network in depicted in the left part of
Fig.~\ref{fig:simple-regnet}.

More formally, a regulatory network is defined as a set $\mathcal{G}$
of regulatory components, each component $G \in \mathcal{G}$ being
associated with a \emph{regulatory function} $K_G$ that provides the
following information:
\begin{itemize}
\item its co-domain defines the range of values $G$ can assume, the
  current value of $G$ being denoted as $x_G$;
\item its arguments define the regulatory components $G$ depends on;
\item its evaluation defines the level toward which $G$ is called to
  evolve, by steps of $\pm 1$.
\end{itemize}
For instance, the right part of Fig.~\ref{fig:simple-regnet}
provides the formal definition of the regulatory network depicted in
the left part. This example is a Boolean network in which all the
components may have values in $\{0,1\}$. In order to model systems
where different thresholds have different influences, more values may
be used for component ranges.

A \emph{state} of a regulatory network is a $\mathcal{G}$-indexed
vector that provides for each component $G\in\mathcal{G}$ its current
level $x_G$. Given a state $s$, and a component $G$, it is possible to
evaluate $K_G$, yielding a target value $x'_G$ for $G$. If $x'_G \neq
x_G$, this defines a possible \emph{evolution} of the system to a
state $s'$ that is such that $s'[H] \defeq s[H]$ for all $H \in
\mathcal{G} \setminus \{G\}$ and:
$$s'[G] \defeq x_G \diff x'_G \defeq 
\left\{\begin{array}{@{}l@{\quad}l}
  x_G + 1 & \mbox{if }x'_G > x_G \mbox{ ,}\\[5pt]
  x_G - 1 & \mbox{if }x'_G < x_G \mbox{ ,}\\[5pt]
  x_G     & \mbox{otherwise.}
\end{array}\right.$$
(Notation $\diff$ will be useful later on to define the Petri nets
semantics.) Such an evolution is denoted by $s \to s'$. Given an
initial state $s_0$ of a regulatory network, it is then possible to
define the \emph{reachable state space} as the smallest graph such
that $s_0$ is a node and, whenever $s$ is a node and $s \to s'$, then
$s'$ is also a node and there is an arc from $s$ toward $s'$. Such a
graph defines a \emph{transition system} that is suitable to perform
model-checking of various reachability- or causality-related
properties.

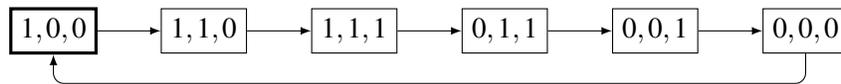
\begin{figure}[b]
\begin{center}
\begin{tikzpicture}[xscale=2,yscale=.7]
\node[draw,very thick] (100) at (0,0) {$1,0,0$};
\node[draw] (110) at (1,0) {$1,1,0$};
\node[draw] (111) at (2,0) {$1,1,1$};
\node[draw] (011) at (3,0) {$0,1,1$};
\node[draw] (001) at (4,0) {$0,0,1$};
\node[draw] (000) at (5,0) {$0,0,0$};
\draw[->] (100) -- (110);
\draw[->] (110) -- (111);
\draw[->] (111) -- (011);
\draw[->] (011) -- (001);
\draw[->] (001) -- (000);
\draw[rounded corners,->] (000) -- +(0,-1) -| (100);
\end{tikzpicture}
\end{center}
\caption{The state space of the regulatory network from
  figure~\ref{fig:simple-regnet} reachable from state $[x_A,x_B,x_C] =
  [1,0,0]$ (depicted in bold).}
\label{fig:state-space}
\end{figure}

For instance, consider state $[1,0,0]$ of the previous network
(indexed as $[x_A,x_B,x_C]$). Only component $B$ may evolve, because
$K_B(x_A) \mapsto 1$, $K_A(x_C) \mapsto 1$ and $K_C(x_B) \mapsto 0$ in
state $[1,0,0]$; this yields a new state $[1,1,0]$. The state space
reachable from $[1,1,0]$ is depicted in Fig.~\ref{fig:state-space}.

\subsection{Modular extension}

In~\cite{ckp10,jb09}, a modular extension of logical regulatory
networks is introduced in order to model regulatory networks
encompassing several cells, each cell being modelled as a
\emph{regulatory module}. This extension consists of two parts:
\begin{itemize}
\item a regulatory module is defined as a regulatory network equipped
      with \emph{inputs} and whose regulatory functions are adapted in
      order to take inputs into account;

\item identified regulatory modules can then be connected, their
      spatial relationships being defined by a map $\delta$
      associating a real value in the segment $[0;1]$ to every pair of
      module identifiers, where value~$0$ represents no influence of
      one module over another, and~$1$ represents a maximal influence.
      We call this value the \emph{neighboring} between two modules,
      and the connected modules form a \emph{regulatory bundle}.
\end{itemize}

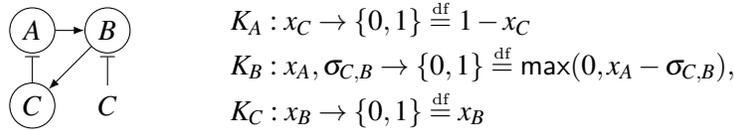
\begin{figure}
\begin{center}
\begin{tikzpicture}
\node[place] (A) at (0,1) {$A$};
\node[place] (B) at (1,1) {$B$};
\node[place] (C) at (0,0) {$C$};
\node (xC) at (1,0) {$C$};
\draw[->](A)--(B);
\draw[->](B)--(C);
\draw[-|,shorten >=1pt](C)--(A);
\draw[-|,shorten >=1pt](xC)--(B);
\node[right] at (2.5,1.1) {$K_A: x_C \to \{0,1\} \defeq 1 - x_C$};
\node[right] at (2.5,.5) {$K_B: x_A,\sigma_{C,B} \to \{0,1\} \defeq
  \max(0, x_A-\sigma_{C,B}),$};
\node[right] at (2.5,-.1) {$K_C: x_B \to \{0,1\} \defeq x_B$};
\end{tikzpicture}
\end{center}
\caption{A regulatory module based on the regulatory network of
  figure~\ref{fig:simple-regnet}.}
\label{fig:module}
\end{figure}

For instance, Fig.~\ref{fig:module} depicts a regulatory module
based on the previous example. It specifies that component $B$ is
repressed by the presence of $C$ outside the module. The uncircled
node thus denotes an input of the module. Notice that no output needs
to be specified since any component can be an input for any neighbor
module. The regulation function $K_B$ is adjusted in order to take
into account the additional information about external $C$. It thus
takes a new argument $\sigma_{C,B}$, called an \emph{integration
  function}, whose unique argument is a set of pairs $(j,x_{C,j})$,
where $j$ is the identifier of a neighbor module and $x_{C,j}$ is the
level of $C$ in module $j$. Intuitively, for every module at non-zero
neighboring to the current module, such a pair belongs to the
argument of $\sigma_{C,B}$. This function is then responsible for
computing a unique value within the admissible range of $C$, that
integrates all the information from the neighborhood. The rationale
behind the use of a $[0;1]$ map to define the neighboring is to take
into account a notion of distance. For instance, diffusion of
chemicals depends on a notion of distance and cannot be easily handled
through the graph of neighborhood: two cells $i$ and $j$ can be in
contact with a cell $k$, but because of their different sizes, the
time used by a chemical diffusing from $i$ to $k$ will be different
from the time taken for a chemical diffusing from $j$ to $k$.

In the example, $K_B$ evaluates to $0$ when $x_A=0$ or $\sigma_{C,B}
\mapsto 1$ and to $1$ otherwise. Function $\sigma_{C,B}$ has not been
specified, for instance, it may be defined as
$$\sigma_{C,B} : (j,x_{C,j})_{j\geq1} \mapsto
  \max_{j\geq1} (\round(\delta(i,j) \cdot x_{C,j})) \mbox{ ,}$$
where $\round$ returns the rounded value of its argument and $i$ is
the identifier of the evolving module. If $j=0$, we define
$\sigma_{C,B} \mapsto 0$ because there is no $C$ in the neighborhood.

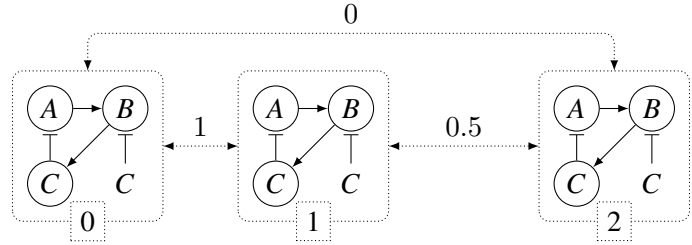
\begin{figure}
\begin{center}
\begin{tikzpicture}
\node[place] (A0) at (0,1) {$A$};
\node[place] (B0) at (1,1) {$B$};
\node[place] (C0) at (0,0) {$C$};
\node (xC0) at (1,0) {$C$};
\draw[->](A0)--(B0);
\draw[->](B0)--(C0);
\draw[-|,shorten >=1pt](C0)--(A0);
\draw[-|,shorten >=1pt](xC0)--(B0);
\draw[rounded corners,densely dotted] (-.5,-.5) rectangle (1.5,1.5);
\node[draw,densely dotted,fill=white] at (.5,-.5) {$0$};
\begin{scope}[shift={(3,0)}]
\node[place] (A1) at (0,1) {$A$};
\node[place] (B1) at (1,1) {$B$};
\node[place] (C1) at (0,0) {$C$};
\node (xC1) at (1,0) {$C$};
\draw[->](A1)--(B1);
\draw[->](B1)--(C1);
\draw[-|,shorten >=1pt](C1)--(A1);
\draw[-|,shorten >=1pt](xC1)--(B1);
\draw[rounded corners,densely dotted] (-.5,-.5) rectangle (1.5,1.5);
\node[draw,densely dotted,fill=white] at (.5,-.5) {$1$};
\end{scope}
\begin{scope}[shift={(7,0)}]
\node[place] (A2) at (0,1) {$A$};
\node[place] (B2) at (1,1) {$B$};
\node[place] (C2) at (0,0) {$C$};
\node (xC2) at (1,0) {$C$};
\draw[->](A2)--(B2);
\draw[->](B2)--(C2);
\draw[-|,shorten >=1pt](C2)--(A2);
\draw[-|,shorten >=1pt](xC2)--(B2);
\draw[rounded corners,densely dotted] (-.5,-.5) rectangle (1.5,1.5);
\node[draw,densely dotted,fill=white] at (.5,-.5) {$2$};
\end{scope}
\draw[<->,densely dotted](1.5,.5)-- node[above] {1} (2.5,.5);
\draw[<->,densely dotted](4.5,.5)-- node[above] {0.5} (6.5,.5);
\draw[<->,densely dotted,rounded corners](.5,1.5)-- (.5,2) --
  node[above] {0} (7.5,2) -- (7.5,1.5);
\end{tikzpicture}
\end{center}
\caption{A bundle of regulatory modules encompassing three copies of
  the module from Fig.~\ref{fig:module}. The neighboring relation
  is depicted by the dotted double-arrowed arcs.}
\label{fig:collect}
\end{figure}

A bundle of such modules, each identified by a unique value (for
instance an integer), can then be formed just by defining the
neighboring relation. For instance, Fig.~\ref{fig:collect} shows
such a bundle. In this example, the effective arguments of
$\sigma_{C,B}$ in module $0$ will be the singleton
$\{(1,x_{C,1})\}$, while for module $1$, it will be the 2-elements set
$\{(0,x_{C,0}), (0.5,x_{C,2})\}$.

A state of a regulatory bundle is also defined as a vector of
component levels, but that is indexed with the components as well as
the module identifiers. For instance, states of the bundle in
Fig.~\ref{fig:collect} may be indexed by $[x_{A,0}, x_{B,0},
  x_{C,0}, x_{A,1}, x_{B,1}, x_{C,1}, x_{A,2}, x_{B,2}, x_{C,2}]$. The
state space of bundles is computed similarly to the state space of
regulatory networks, except that it involves the evaluation of
integration functions, taking into account the neighboring
relation~$\delta$.

\subsection{Petri net semantics}

In order to analyze regulatory bundles, a Petri net semantics has
been defined~\cite{ckp10}. For this purpose, a colored variant of
Petri nets~\cite{cpn} has been used in which:
\begin{itemize}
\item \emph{places} model containers of \emph{tokens}, the latter
  being arbitrary values, \eg, pairs of integers as in the following;
\item \emph{transitions} model activities that consume and produce
  tokens in places;
\item \emph{arcs} connect places to transitions (and vice-versa),
  indicating how tokens are consumed or produced by the transitions.
\end{itemize}

\begin{figure}
\begin{center}
\begin{tikzpicture}[xscale=6,yscale=2]
\node[place] (A) at (1,2) {};
\node[above] at (A.north) {$A$};
\node[place] (B) at (2,1) {};
\node[above] at (B.north) {$B$};
\node[place] (C) at (1,0) {};
\node[below] at (C.south) {$C$};
\node[transition] (tA) at (0,2) {$t_A$};
\draw[->](node cs:name=A,angle=170) -- node[above] {$(i,x_{A,i})$}
  (node cs:name=tA,angle=10);
\draw[->](node cs:name=tA,angle=-10) -- node[below] {$(i,x_{A_i} \diff
  K_A(x_{C,i}))$} (node cs:name=A,angle=190);
\draw[<->,rounded corners,near end](tA) |- node[below] {$(i,x_{C,i})$}
  (C);
\node[transition] (tC) at (2,0) {$t_C$};
\draw[->](node cs:name=C,angle=10) -- node[above] {$(i,x_{C,i})$}
  (node cs:name=tC,angle=170);
\draw[->](node cs:name=tC,angle=190) -- node[below] {$(i,x_{C,i} \diff
  K_C(x_{B,i}))$} (node cs:name=C,angle=-10);
\draw[<->](tC) -- node[right] {$(i,x_{B,i})$} (B);
\node[transition] (tB) at (1,1) {$t_B$};
\draw[->](node cs:name=B,angle=170) -- node[above] {$(i,x_{B,i})$}
  (node cs:name=tB,angle=10);
\draw[->](node cs:name=tB,angle=-10) -- node[below]
    {$(i,x_{B,i} \diff K_B(x_{A,i},\sigma_{C,B}(\vec{x_C})))$} 
  (node cs:name=B,angle=190);
\draw[<->](C) -- node[left] {$\vec{x_C} \defeq \{(j,x_{C,j}) \mid
  \delta(i,j) > 0\}$} (tB);
\draw[<->](A) -- node[right] {$(i,x_{A,i})$} (tB);
\end{tikzpicture}
\end{center}
\caption{The Petri net defining the semantics of the regulatory
  bundle from figure~\ref{fig:collect}.}
\label{fig:pn}
\end{figure}
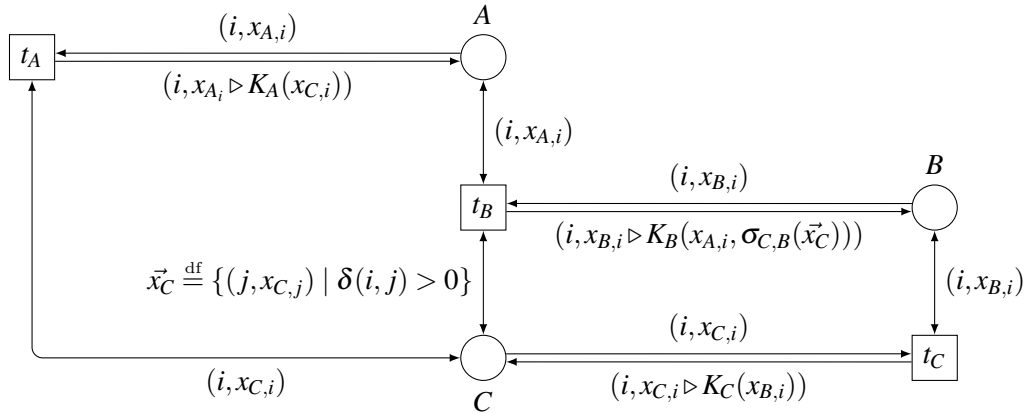

Fig.~\ref{fig:pn} shows an example of such a Petri net, that is
actually the semantics of the regulatory bundle from
Fig.~\ref{fig:collect}. As usual, places are depicted by round
nodes, transitions by square nodes and arcs by directed edges. In this
net, places $A$, $B$ and $C$ are used to store the current level of
each component in each module. For this purpose, they may be
\emph{marked} with tokens that are pairs $(i,x_{G,i})$ storing the
level of a component $G$ for a module identified by $i$. The marking
thus corresponds to the state of the regulatory bundle. Then,
transitions are used to implement the evolution of components.
Consider for instance transition $t_A$, it consumes a pair
$(i,x_{A,i})$ from place $A$, as well as a pair $(i,x_{C,i})$ from
$C$; the latter is reproduced in the same place, as denoted by the
double-arrowed arc, while the former is replaced by a pair $(i,x_{A,i}
\diff K_A(x_{C,i}))$, that is, the evolution of $A$ in module $i$.
This evolution may happen for any module identifier, the consistency
of consumed/produced values is ensured by the selection of a unique
value for $i$ during the execution, or \emph{firing}, of $t_A$.
Transition $t_C$ behaves similarly. Transition $t_B$ is slightly
different because the evolution of $B$ depends on $A$ in the evolving
module but also on the values of $C$ in its neighborhood. So, $t_B$
consumes $(i,x_{A,i})$ and $(i,x_{B,i})$ as expected, but also a set
of pairs $(j,x_{C,j})$ that is computed dynamically on the arc from
$C$ to $t_B$, with respect to the function $\delta$ that encodes the
neighboring relation. This set of pairs is bound to a variable
$\vec{x_C}$ that is then used to compute $\sigma_{C,B}$ during the
evaluation of $K_B$, allowing to produce in place~$B$ the new level of
component~$B$ for module~$i$.

The initial state of the regulatory bundle is rendered as the
initial marking of the Petri net that defines its semantics: for each
component $G$ in a module $i$ with the initial level $x_{G,i}$, one
token $(i,x_{G,i})$ is added to the marking of place $G$.

\subsection{Topological collections}

We have just seen that the neighboring relation within a regulatory
bundle can be encoded as a function $\delta$ that returns for each
pair $(i,j)$ of module identifiers (with $i \neq j$), the neighboring
value of $i$ with respect to $j$. This is a very general encoding but
it is not related to any kind of geometrical, spatial or topological
information. Our goal in this paper is to provide such an information
in a way that allows for: spatial evolutions, enumerations of the
possible evolutions, graphical rendering of spatial information. To do
so, we resort to \emph{topological collections}.

Topological collections have been introduced in~\cite{giavitto02} to
describe arbitrary complex spatial structures that appear in
biological systems~\cite{giavitto03bis}. They have been used to
represents states of dynamical systems with a time varying
structure~\cite{Giavitto03,Giavitto2008}. We will use them in this
paper as a unifying framework to represent arbitrary neighborhood
relationships. In this context, a neighborhood relationship between
modules is represented by a labelled graph. Each vertex in this graph
is labelled by a module identifier and a given identifier is the label
of only one vertex. Two modules $i$ and $j$ are neighbors and
$\delta(i,j)= \alpha$ if there is an edge labelled by $\alpha$ between
the vertices associated with $i$ and $j$. However, as we will see later
on, such graphs will not be represented explicitly. Instead, relation
$\delta$ will be computed dynamically with respect to some lower-level
neighboring information.

Topological collections of various kinds have been implemented in an
experimental programming languages called MGS~\cite{mgs}. This
language is used here to implement a database that records the
neighborhood relationships and that can be queried and updated
efficiently.

\section{Representing spatial information}
\label{sec:spatial}

In this section, we apply topological collections to the
representation of spatial information in two ways. A first solution
based on predefined grids is proposed: it is simple and efficient, but
it is limited to represent systems in which modules may be inserted or
moved only into existing ``holes'' in the grid. In order to overcome
this limitation, we then propose another, more complex solution based
on undirected graphs with limited degrees.

\subsection{Placing modules on grids}
\label{sec:gbf}

Simple topological collections can be defined as \emph{group based
  fields} (GBF), that can be considered as associative arrays whose
indexes are elements in a group~\cite{giavitto01c}. The latter is
defined by a \emph{finite presentation}: a set of generators together
with some constraints on their combinations. Thus a GBF can be
pictured as a labelled graph where the underlying graph is the Cayley
graph of the finite presentation. The labels are the values
associated with the vertices and the generators are associated with the
edges.

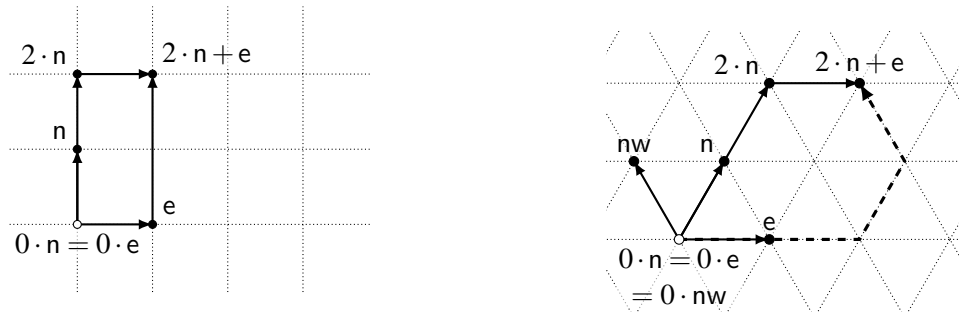
\begin{figure}
\begin{center}
\begin{tikzpicture}
\draw[densely dotted] (-.9,-.9) grid (3.9,2.9);
\draw[fill] (1,0) circle (1.5pt) node[above right]
     {\GBF{e}};
\draw[fill] (0,2) circle (1.5pt) node[above left]
     {$2\cdot\GBF{n}$};
\draw[fill] (0,1) circle (1.5pt) node[above left]
     {\GBF{n}};
\draw[fill] (1,2) circle (1.5pt) node[above right,cover]
     {$2\cdot\GBF{n}+\GBF{e}$};
\draw[thick,->] (0,0) -- (1,0);
\draw[thick,->] (0,0) -- (0,2);
\draw[thick,->] (0,0) -- (0,1);
\draw[thick,->] (1,0) -- (1,2);
\draw[thick,->] (0,2) -- (1,2);
\draw[fill=white] (0,0) circle (1.5pt) node[below,cover]
     {$0\cdot\GBF{n} = 0\cdot\GBF{e}$};
\begin{scope}[shift={(8,-.2)},scale=1.2]
\path[clip](-.8,-.8) rectangle (3.1,2.3);
\foreach \x in {-2,...,3}
  \foreach \y in {-2,...,3}
    \draw[densely dotted] (0,0) ++(0:\x) ++(60:\y) -- ++(60:1)
    -- ++(-60:1) -- +(180:1);
\draw[fill] (0,0) ++(60:1) circle (1.5pt) node[above left,cover]
  {\GBF{n}};
\draw[thick,->] (0,0) -- ++(60:1);
\draw[fill] (0,0) ++(120:1) circle (1.5pt) node[above,cover]
  {\GBF{nw}};
\draw[thick,->] (0,0) -- ++(120:1);
\draw[fill] (0,0) ++(60:2) circle (1.5pt) node[above left,cover]
  {$2\cdot\GBF{n}$};
\draw[thick,->] (0,0) -- ++(60:2);
\draw[fill] (0,0) ++(0:1) circle (1.5pt) node[above,cover]
  {\GBF{e}};
\draw[thick,->] (0,0) -- ++(0:1);
\draw[fill] (0,0) ++(60:2) ++(0:1) circle (1.5pt) node[above,cover]
  {$2\cdot\GBF{n}+\GBF{e}$};
\draw[thick,->] (0,0) ++(60:2) -- ++(0:1);
\draw[very thick,dashed,->] (0,0) -- ++(0:2) -- ++(60:1) -- ++(120:1);
\draw[fill=white] (0,0) circle (1.5pt) node[below,cover,text
  width=16mm,text centered] {$0\cdot\GBF{n} = 0\cdot\GBF{e}$\newline
  $= 0\cdot\GBF{nw}$};
\end{scope}
\end{tikzpicture}
\end{center}
\caption{Left: a GBF defining a square grid, with two generators
  \GBF{e} and \GBF{n}. Right: a GBF defining a triangular grid with
  three generators \GBF{e}, \GBF{n} and \GBF{nw}, and a constraint
  $\GBF{n} - \GBF{nw} = \GBF{e}$.}
\label{fig:grids}
\end{figure}

For instance, in order to define a square grid, we may use two
generators \GBF{e} (east) and \GBF{n} (north), supporting addition,
difference and multiplication by an integer. This is illustrated in
the left part of Fig.~\ref{fig:grids}.

Similarly, a triangular grid can be defined by means of three
generators \GBF{n}, \GBF{e} and \GBF{nw} (north-west) and a constraint
$\GBF{n} - \GBF{nw} = \GBF{e}$, as illustrated in the right part of
Fig.~\ref{fig:grids}. Notice that such a triangular grid is adequate
to represent cells with a hexagonal shape, since the grid can be paved
with hexagons centered at the positions in the grid. As shown by the
dashed path, we have $2\cdot\GBF{n}+\GBF{e} = 2\cdot\GBF{e} + \GBF{n}
+ \GBF{nw}$, which can be also checked in an algebraic way, by
substituting \GBF{nw} with $\GBF{n} - \GBF{e}$ in this equality as
allowed by the constraint.

The GBF structure is thus adequate to define the arrangement on a
grid, in any number of dimensions. In such grids, a distance can be
naturally defined as the minimum number of steps in order to reach one
point from the other (this is the approach of \emph{geometric group
  theory}). For instance, in the triangular grid of
Fig.~\ref{fig:grids}, points at \GBF{e} and \GBF{n} are at distance
1 because only one step in direction \GBF{nw} is required to reach the
latter from the former; similarly, points \GBF{n} and
$2\cdot\GBF{n}+\GBF{e}$ are at distance 2. Let us denote by
$\Delta(x,y)$ the distance between two points $x$ and $y$. It is easy
to check that $\Delta(x,y) = \Delta(y,x)$ for any $x$ and $y$.

Such a distance can be used to implement our neighboring relation,
for instance, for $x \neq y$, we could define:
$$\delta(x,y) \defeq {1 \over \Delta(x,y)}$$
which matches the intuition that $\delta(x,y) = 1$ for two immediate
neighbors while $\delta(x,y)$ converges toward $0$ when $x$ and $y$
become farther one each other.

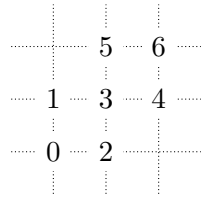
\begin{figure}
\begin{center}
\begin{tikzpicture}[scale=.7]
\draw[densely dotted] (-.8,-.8) grid (2.8,2.8);
\node[fill=white] at (0,0) {0};
\node[fill=white] at (0,1) {1};
\node[fill=white] at (1,0) {2};
\node[fill=white] at (1,1) {3};
\node[fill=white] at (2,1) {4};
\node[fill=white] at (1,2) {5};
\node[fill=white] at (2,2) {6};
\end{tikzpicture}
\end{center}
\caption{A square grid with modules (denoted by their identifiers)
  placed on it.}
\label{fig:limitation}
\end{figure}

The main drawback with grids is that inserting new elements is
possible only if a hole is already present. Consider for instance
Fig.~\ref{fig:limitation} and assume that the modules represent
cells forming a tissue. It would be difficult to model the division of
cell~3, because there is no free position adjacent to~3. But in many
biological system, we may expect that the division of cell~3 results
in ``pulling away'' the neighboring cells. Similarly, if cell~3 is
called to die, removing it from the grid will result in a disconnected
tissue, which may be undesirable too.

However, the simplicity of grids is well adapted to model, \eg,
accretive growth that occurs at the borders of a tissue like leaves in
plants~\cite{GiavittoMC02}. Notice also that the graphical rendering
of GBF is trivial to obtain because the grids are predefined.

\subsection{A generalization of grids}

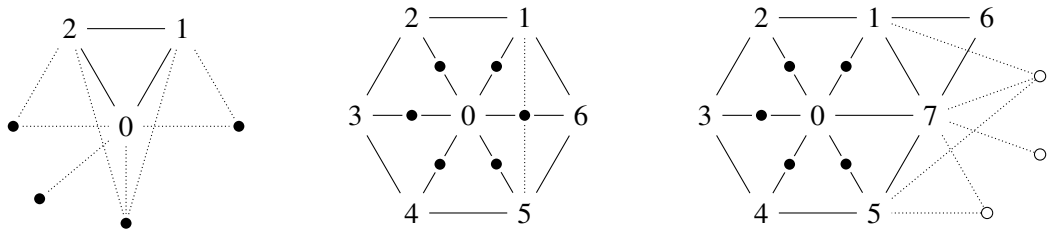
\begin{figure}[b]
\begin{center}
\begin{tikzpicture}[scale=1.5]
\node at (0:0) (N0) {$0$};
\node at (60:1) (N1) {$1$};
\node at (120:1) (N2) {$2$};
\node[draw=white,line width=1.5pt,circle,fill,inner sep=2pt] at
  (180:1) (Na) {};
\node[draw=white,line width=1.5pt,circle,fill,inner sep=2pt] at (0:1)
  (Nb) {}; 
\node[draw=white,line width=1.5pt,circle,fill,inner sep=2pt] at
  (220:1) (Nc) {}; 
\node[draw=white,line width=1.5pt,circle,fill,inner sep=2pt] at
  (270:.86) (Nd) {};
\draw (N0)--(N1)--(N2)--(N0);
\draw[densely dotted] (N2) -- (Na) -- (N0) -- (Nb) -- (N1) (N0) --
 (Nc) (N0) -- (Nd) -- (N1) (Nd) -- (N2);
\end{tikzpicture}
\hfil
\begin{tikzpicture}[scale=1.5]
\node at (0:0) (N0) {$0$};
\foreach \n/\a in {1/60, 2/120, 3/180, 4/240, 5/300, 6/0}
{
  \node at (\a:1) (N\n) {$\n$};
  \draw (N0) -- node[name=M\n,draw=white,line
    width=1.5pt,circle,fill,inner sep=2pt] {} (N\n);
}
\draw (N1) -- (N2) -- (N3) -- (N4) -- (N5) -- (N6) -- (N1);
\draw[densely dotted] (N5) -- (M6) -- (N1);
\end{tikzpicture}
\hfil
\begin{tikzpicture}[scale=1.5]
\node at (0:0) (N0) {$0$};
\foreach \n/\a in {1/60, 2/120, 3/180, 4/240, 5/300}
{
  \node at (\a:1) (N\n) {$\n$};
  \draw (N0) -- node[draw=white,line width=1.5pt,circle,fill,inner
    sep=2pt] {} (N\n);
}
\path ++(0:1) ++(60:1) node (N6) {$6$};
\path ++(0:1) ++(-60:1) node[draw,fill=white,circle,inner sep=1.5pt]
  (N6x) {};
\path (-10:2) node[draw,fill=white,circle,inner sep=1.5pt] (N6y) {};
\path (10:2) node[draw,fill=white,circle,inner sep=1.5pt] (N6z) {};
\node at (0:1) (N7) {$7$};
\draw (N1) -- (N2) -- (N3) -- (N4) -- (N5) -- (N7) -- (N0) (N1) --
  (N6) -- (N7) -- (N1);
\draw[densely dotted] (N6y) -- (N7) -- (N6x) -- (N5)
  (N1) -- (N6z) -- (N5) (N7) -- (N6z);
\end{tikzpicture}
\end{center}
\caption{Left: a 6-BDG with 3 nodes and 4 possible insertion points
  from node~$0$. Middle: a 6-BDG with 7 nodes and 6 possible insertion
  points from node~$0$. Right: the same 6-BDG after the insertion of
  node $7$ on edge $0\mbox{---}6$. Dotted lines represent further
  possibilities to place node~$6$.}
\label{fig:6-graph}
\end{figure}

In order to overcome the limitations of grids in the presence of cells
migration, division or apoptosis, we now consider a topological
collection based on undirected graphs, whose degree will be kept
bounded. Modules will be located on the nodes of such graphs and edges
will represent immediate neighboring. For instance, by bounding
degree to~6, which we call a \emph{6-bounded degree graph} (6-BDG), we
shall obtain a result similar to triangular grids, as illustrated in
Fig.~\ref{fig:6-graph}.

Consider first the left part of Fig.~\ref{fig:6-graph}: it depicts a
6-BDG with three nodes~$0$, $1$ and~$2$, each being an immediate
neighbor of the others, as depicted by the plain lines. The black
dots and the dotted lines depict all the free locations next to
node~$0$. Indeed, a new immediate neighbor next to~$0$ may be also a
neighbor of~$1$, $2$, both, or none of them. Notice that the
geometrical position of nodes is not relevant, but only the edges
between them are important. For instance, the black dot at the bottom
may be equivalently depicted inside the triangle formed by nodes $0$,
$1$ and $2$. As long as node degrees is limited, the number of free
locations next to a given node is limited as well.

When no new node can be inserted without violating the degree
boundary, we proceed by splitting an existing edge. This is the case
for instance in the middle of Fig.~\ref{fig:6-graph} where the black
dots depict all the possible insertion points of a node next to~$0$.
Dotted lines indicate the neighboring that would result from
inserting a node on the edge from~$0$ to~$6$. This insertion results
in ``pulling away'' node~$6$, which is depicted in the right part of
the figure. White dots and dotted lines indicate alternate positions
for the pulled node~$6$. In every case, there is also a limited number
of possible graph transformations.

Deletion of nodes is handled in the most simple way, just by removing
them in the graph. How the neighbors of a deleted node are then
reconnected is modelled as a migration process. For instance, consider
again the left part of Fig.~\ref{fig:6-graph}, in particular the two
left-most black dots, and assume a cell at the position of the lower
dot (the one only connected to $0$). This cell may be isolated
from~$2$ because another cell, placed at the second dot, has just
disappeared. The remaining cell may then migrate to fill this freshly
freed position, or any of the positions depicted as black dots. In
other words, a cell's death creates a hole that may then be filled by
the migration of other cells around.

This system is thus much more flexible than the previous solution, but
at the price of more complexity. However, this complexity is largely
alleviated by MGS that provides almost ready-made solutions to
implement such a system. An additional difficulty is to produce a
graphical rendering of an $n$-BDG. Here also, MGS simplifies the
problem by proposing graph layout algorithms that have been already
proved relevant to represent graphically cells
positions~\cite{Reuille2006,iccs05}.

In an $n$-BDG, a notion of distance can be defined as the length of
the shortest path between two nodes. If such a path does not exist,
\ie, if the graph is not connected, the distance can be assumed
infinite. Using this distance, we can then define the neighboring
relation $\delta$ exactly as in GBFs.

\section{Spatialized regulatory bundles}
\label{sec:contrib}

This section introduces an extension of regulatory bundles in
order to take into account the spatial representations we have defined
above. As a first step, we define a uniform interface for spatial
information, that is independent of the actually chosen
representation. The idea is that, at the level of semantics, this
interface is queried from the Petri net part and, on the other hand,
implemented on the top of a topological collection.

\subsection{Spatial interfaces}

Let us denote by $\mathbb{I}$ the set of module identifiers. We also
define a set $\mathbb{L}$ of (abstract) \emph{locations} that
represent the positions in the chosen spatial representation. A
\emph{spatial interface} is a tuple $(\theta, \delta, \eta)$ where:
\begin{itemize}
\item $\theta$ is a partial function $\mathbb{I} \to \mathbb{L}$ that
  maps some identifiers to locations, its domain being denoted by
  $\mathbb{I}_\theta$;
\item $\delta : (\mathbb{I}_\theta)^2 \setminus \{(i,i) \mid i \in
  \mathbb{I}_\theta\} \to [0;1]$ is the \emph{neighboring relation}
  as previously;
\item $\eta : \mathbb{I}_\theta \to 2^\mathbb{L}$ returns for every
  module a set of \emph{empty locations} in the immediate
  neighborhood of the module (this set may be empty).
\end{itemize}
In order to simplify notations, we may consider these functions as
sets of pairs.

Intuitively, $\mathbb{I}_\theta$ defines the \emph{allocated
  identifiers}, \ie, those that correspond to actually existing
modules, or living cells. On this allocated identifiers, $\delta$
encodes the neighboring relation. Finally, $\eta$ is used when a cell
is called to migrate or divide: in both cases, a new location next to
the cell is needed, either as a new location for a migrating cell, or
as a location for the newly created cell. Apoptosis, or freeing of a
location after a migration, are simply modelled by restricting the
domain of $\theta$.

When a GBF-based representation is used, $\mathbb{L}$ is simply the
set of all possible normalized linear combinations of generators,
$\delta$ can be implemented as shown in section~\ref{sec:gbf}, and
$\eta(i) \defeq \{\ell \in \mathbb{L} \setminus
\theta(\mathbb{I}_\theta) \mid \Delta(\theta(i),\ell) = 1\}$, \ie, all
the locations not already allocated and at distance~1 from the
location of~$i$. Exactly the same approach can be used for $n$-BDG,
except that $\theta$ in the case must record positions is the graph,
which is even simpler if the nodes are numbered consistently with the
modules.

\subsection{Regulatory modules with transformations}

So far, a regulatory module defines the rules to change the levels of
its components. This is not enough to model the migration, division or
apoptosis processes. So we extend regulatory modules with three
functions to specify these transformations:
\begin{itemize}
\item a function $K_{\dag}$ for apoptosis, whose arguments are as for
  the regulatory functions, \ie, levels of components within the
  module or its neighborhood. It returns a Boolean value indicating
  whether the cell modelled by the module should die or not;
\item a function $K_{\leadsto}$ for migration, whose arguments are as
  for $K_{\dag}$, plus an additional $L \subseteq \mathbb{L}$. It
  returns a subset of $L$ that indicates the locations where the cell
  is allowed to migrate, taking into account information about the
  local and external levels of regulatory components;
\item a function $K_{\%}$ for division, whose arguments are as for
  $K_{\leadsto}$ and that returns the set of locations where the new
  cell resulting from the division can be placed. In this paper, we
  assume that when a cell $i$ divides, it yields the same cell $i$
  together with a new, identical cell associated with an unallocated
  identifier $j \in \mathbb{I} \setminus \mathbb{I}_\theta$. As shown
  later on, this policy can be easily replaced by another that is more
  suited to the biological system of interest.
\end{itemize}

\begin{figure}
\begin{center}
\begin{tikzpicture}
\node[place] (A) at (0,1) {$A$};
\node[place] (B) at (1,1) {$B$};
\node[place] (C) at (0,0) {$C$};
\node (xC) at (1,0) {$C$};
\draw[->](A)--(B);
\draw[->](B)--(C);
\draw[-|,shorten >=1pt](C)--(A);
\draw[-|,shorten >=1pt](xC)--(B);
\node[place,thick,densely dotted] (die) at (-1,1) {$\dag$};
\draw[-|,shorten >=1pt](A)--(die);
\node (xA) at (-2,.5) {$A$};
\draw[-|,shorten >=1pt](xA)--(die);
\node[place,thick,densely dotted] (split) at (-1,0) {$\%$};
\draw[-|,shorten >=1pt](A)--(split);
\draw[->](xA)--(split);
\node[place,thick,densely dotted] (move) at (.5,2) {$\leadsto$};
\draw[->](A)--(move);
\draw[->](B)--(move);
\node[right] at (2.5,1.5) {$K_{\dag}: x_A, \sigma_{A,\dag} \to x_A = 0
  \land \sigma_{A,\dag} = 0$};
\node[right] at (2.5,.8) {$K_{\leadsto}: x_A, x_B, L \to L
  \mbox{ if } x_A + x_B > 0 \mbox{ else } \emptyset$};
\node[right] at (2.5,.1) {$K_{\%}: x_A,\sigma_{A,\%}, L \to L
  \mbox{ if } \sigma_{A,\%} - x_A > 0 \mbox{ else } \emptyset$};
\end{tikzpicture}
\end{center}
\caption{A regulatory module with transformations, based on the
  regulatory module of Fig.~\ref{fig:module}. Special nodes allow to
  depict how apoptosis ($\dag$), migration ($\leadsto$) and division
  ($\%$) are regulated. Function $\sigma_{A,\dag}$ and $\sigma_{A,\%}$
  may be defined for instance as $\sigma_{C,B}$ previously.}
\label{fig:extended-module}
\end{figure}
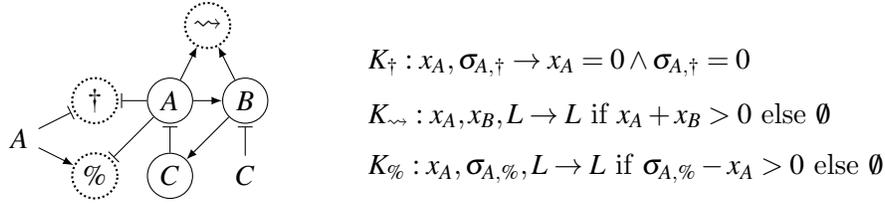

How these transformation processes are regulated can be graphically
rendered as special nodes, similar to regulatory components nodes. See
for instance Fig.~\ref{fig:extended-module} that defines the
transformations (right part) and depict them (left part):
\begin{itemize}
\item the cell may die whenever there is no more $A$ expressed within
  the cell or its environment, which indeed means that $A$ is a
  repressor for apoptosis;
\item the cell may migrate to any available location if (and only if)
  there is some $A$ or $B$ expressed within the cell;
\item the cell may divide, spawning to any available location, if (and
  only if) there is some $A$ expressed in its environment but not
  inside the cell.
\end{itemize}
Notice that these new nodes have been depicted dotted because they can
be regulated but cannot be used as regulators themselves.

Notice that we defined $K_{\leadsto}$ and $K_{\%}$ so that they can
return a subset of their argument $L$. This has not been exploited in
the example to keep it simple. But it is possible to implement complex
functions that select among the locations in $L$ only the ``best''
ones for migration or division. For instance, a function
$K_{\leadsto}$ with an argument $\sigma_{G,\leadsto}$ may be designed
to migrate in the directions of cells with a given level of component
$G$ among those proposed in its argument~$L$.

\subsection{Spatialized regulatory bundles}

A \emph{spatialized regulatory bundle} can be defined as a set of
regulatory modules with transformations, together with a spatial
interface. For example, we could build a bundle similar to that of
Fig.~\ref{fig:collect} by using the module with transformations
defined in Fig.~\ref{fig:extended-module} and equipping it with the
following GBF-based spatial interface:
\begin{itemize}
\item $\theta \defeq \{(0,0\cdot\GBF{n}), (1,\GBF{e}),
  (2,3\cdot\GBF{e})\}$;
\item $\delta(i,j) \defeq 1/\Delta(\theta(i),\theta(j)) \mbox{ if }
  \Delta(\theta(i),\theta(j)) \leq 2 \mbox{ else } 0$;
\item $\eta(i) \defeq \{\ell \in \mathbb{L} \setminus
  \theta(\mathbb{I}_\theta) \mid \Delta(\theta(i),\ell) = 1\}$ as
  previously defined.
\end{itemize}
One can check that $\delta$ returns the same values as in the previous
version of our example. (And this actually holds independently of
whether square or triangular grids are used.)

The Petri net semantics is defined as before except that we need to
add transitions to implement the spatial transformations. Moreover,
the Petri net evolution is dependent on the spatial interface, that in
turn may be modified when spatial transformations are executed. In the
following, we consider that the interface elements are stored into
some kind of global variables, that may be accessed from any part of
the Petri net. (This could be implemented directly within Petri nets
but would result in more complex notations.)

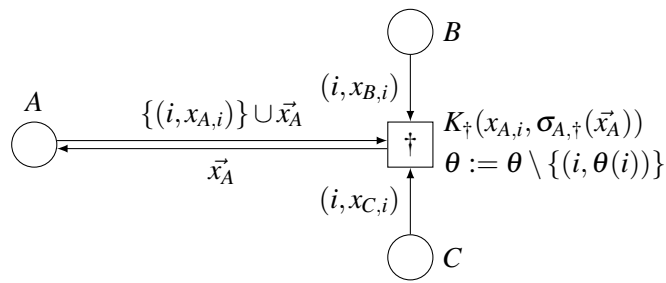
\begin{figure}[b]
\begin{center}
\begin{tikzpicture}[xscale=5,yscale=1.5]
\node[place] (A) at (0,0) {};
\node[above] at (A.north) {$A$};
\node[transition] (die) at (1,0) {$\dag$};
\node[right] at (die.east) {$\begin{array}{@{}l}
  K_{\dag} (x_{A,i}, \sigma_{A,\dag}(\vec{x_A})) \\[2pt]
  \theta := \theta \setminus\{(i,\theta(i))\}
  \end{array}$};
\draw[->](node cs:name=A,angle=10) -- node[above] {$\{(i,x_{A,i})\}
  \cup \vec{x_A}$} (node cs:name=die,angle=170);
\draw[->](node cs:name=die,angle=190) -- node[below] {$\vec{x_A}$}
  (node cs:name=A,angle=-10);
\node[place] (B) at (1,1) {};
\node[right] at (B.east) {$B$};
\draw[->](B) -- node[left] {$(i,x_{B,i})$} (die);
\node[place] (C) at (1,-1) {};
\node[right] at (C.east) {$C$};
\draw[->](C) -- node[left] {$(i,x_{C,i})$} (die);
\end{tikzpicture}
\end{center}
\caption{The Petri net semantics of transformation $K_{\dag}$, where
  $\vec{x_A} \defeq \{(j,x_{C,j}) \mid \delta(i,j) > 0\}$.}
\label{fig:pn-die}
\end{figure}

The semantics of the apoptosis transformation for our example is
depicted in Fig.~\ref{fig:pn-die}. It consumes all the current
component levels for a module $i$, plus the levels of $A$ in the
neighborhood of $i$. The various levels of $A$ are used to evaluate
the Boolean expression $K_{\dag} (x_{A,i}, \sigma_{A,\dag}
(\vec{x_A}))$ that appears as the \emph{guard} of the transition, \ie,
a condition for its firing. Whenever the transition fires, it also
updates $\theta$ by removing $i$ from its domain; the transition also
reproduce $\vec{x_A}$ in place $A$. The effect of this evolution is to
remove from the places and from $\theta$ all the information related
to module $i$, which corresponds to how its disappearing is rendered.

\begin{figure}
\begin{center}
\begin{tikzpicture}[yscale=1.5]
\node[transition] (move) at (1,0) {$\leadsto$};
\node[right] at (move.east) {$\begin{array}{@{}l}
  \ell \in K_{\leadsto} (x_{A,i}, x_{B,i}, \eta(\theta(i)))\\[2pt]
  \theta := \{(i,\ell)\} \cup \theta \setminus\{(i,\theta(i))\}
  \end{array}$};
\node[place] (B) at (1,-1) {};
\node[right] at (B.east) {$B$};
\draw[<->](B) -- node[left] {$(i,x_{B,i})$} (move);
\node[place] (A) at (1,1) {};
\node[right] at (A.east) {$A$};
\draw[<->](A) -- node[left] {$(i,x_{A,i})$} (move);
\end{tikzpicture}
\end{center}
\caption{The Petri net semantics of transformation
  $K_{\leadsto}$.}
\label{fig:pn-move}
\end{figure}

The Petri net semantics of the migration transformation for our
example is depicted in Fig.~\ref{fig:pn-move}. As usual, it consumes
and reproduces the necessary levels in the adequate places. Then,
thanks to the guard $\ell \in K_{\leadsto} (x_{A,i}, x_{B,i},
\eta(\theta(i)))$, a target location for migration is selected among
the possible ones, and $\theta$ is updated accordingly upon firing.

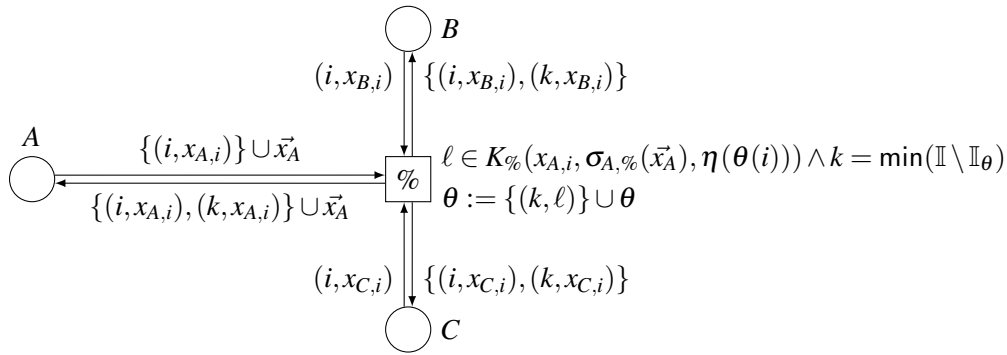
\begin{figure}[b]
\begin{center}
\begin{tikzpicture}[xscale=5,yscale=2]
\node[place] (A) at (0,0) {};
\node[above] at (A.north) {$A$};
\node[transition] (split) at (1,0) {$\%$};
\node[right] at (split.east) {$\begin{array}{@{}l}
  \ell \in K_{\%} (x_{A,i}, \sigma_{A,\%}(\vec{x_A}),
    \eta(\theta(i))) \land k = \min(\mathbb{I} \setminus
    \mathbb{I}_\theta) \\[2pt]
  \theta := \{(k,\ell)\} \cup \theta
  \end{array}$};
\draw[->](node cs:name=A,angle=10) -- node[above] {$\{(i,x_{A,i})\}
  \cup \vec{x_A}$} (node cs:name=split,angle=170);
\draw[->](node cs:name=split,angle=190) -- node[below]
     {$\{(i,x_{A,i}), (k,x_{A,i})\} \cup \vec{x_A}$} (node
     cs:name=A,angle=-10);
\node[place] (B) at (1,1) {};
\node[right] at (B.east) {$B$};
\draw[near start,->](node cs:name=B,angle=-100) -- node[left]
  {$(i,x_{B,i})$} (node cs:name=split,angle=100);
\draw[near end,->](node cs:name=split,angle=80) -- node[right]
  {$\{(i,x_{B,i}), (k,x_{B,i})\}$} (node cs:name=B,angle=-80);
\node[place] (C) at (1,-1) {};
\node[right] at (C.east) {$C$};
\draw[near start,->](node cs:name=C,angle=100) -- node[left]
  {$(i,x_{C,i})$} (node cs:name=split,angle=-100);
\draw[near end,->](node cs:name=split,angle=-80) -- node[right]
  {$\{(i,x_{C,i}), (k,x_{C,i})\}$} (node cs:name=C,angle=80);
\end{tikzpicture}
\end{center}
\caption{The Petri net semantics of transformation $K_{\%}$, where
  $\vec{x_A}$ is the same as previously.}
\label{fig:pn-split}
\end{figure}

Finally, the Petri net semantics of the division transformation for
our example is depicted in Fig.~\ref{fig:pn-split}. It reads
(consume/produce) all the information from a module $i$ and, thanks to
its guard, binds to $\ell$ an acceptable location. The second part of
the guard, $k = \min(\mathbb{I} \setminus \mathbb{I}_\theta)$, allows
to bind to $k$, an unallocated identifier for the newly created cell.
Its component levels are saved to the appropriate places as exact
copies of the levels in module $i$ so that the division creates two
identical cells (at two adjacent locations), as required by the policy
chosen above. If another rule is preferred, it can be implemented
simply by changing the Petri net semantics of $K_{\%}$; for instance
by computing new states and/or locations for the cells resulting from
the division. Notice that we choose for~$k$ the smallest unallocated
identifier (assuming that $\mathbb{I}$ is ordered) in order to reduce
the combinatorial explosion during the evaluation of the guard. If
$\min$ would not be used, every unallocated identifier could be bound
to $k$, leading to as many successor states as the size of $\mathbb{I}
\setminus \mathbb{I}_\theta$. On the contrary, it is important that
the selection of $\ell$ is not constrained in the same way, which
allows to explore all the possible evolutions of the spatial
structure.

\subsection{Implementation and applications}

Spatialized regulatory bundles have been implemented as a prototype,
using SNAKES~\cite{PNN08,snakes} for the Petri net part, and
MGS~\cite{mgs} for the topological collections part. SNAKES is a
general purpose Petri net library implemented in Python, that is
particularly suitable for the quick development of prototypes. This
implementation allows to define regulatory modules as special cases of
Python classes in which methods implement the evolution and
transformation rules. Bundles can then be populated with instances of
these modules and their locations can be specified as MGS expressions,
consistently with the chosen spatial interface. The Petri nets
semantics can be automatically computed (and then exported or drawn)
and its state space can be explored for analysis purpose.

For our implementation, we have actually reused a previous prototype
developed in the context of~\cite{ckp10} and extended it as follows:
\begin{itemize}
\item spatial interfaces have been implemented within MGS on the top
  of both GBFs and BDGs;
\item a bridge to allow calls to MGS from Python has been realized by
  driving the MGS shell from Python through a pair of pipes;
\item the original implementation of the neighboring relation (a
  simple map from pairs of identifiers to float values) has been
  replaced by queries to the spatial interface;
\item the semantics of transformations for cells migration, division
  or apoptosis has been added.
\end{itemize}

This prototype implementation has been applied to the study of two
simple biology-inspired examples, allowing to test various features of
our framework~\cite{castagnet}:
\begin{enumerate}
\item A system with transcription signal cascading resulting in a cell
  division allowed to observe the alternation of growth phases and
  divisions, which can be interpreted as a mutual influence between
  components evolutions and spatial transformations. Moreover, by
  controlling a chosen set of growth factors, the duplication of the
  modelled cells could be controlled. This example used a GBF
  representation for the spatial information.
\item A simplified model of epithelial cellular healing has been
  defined with cells arranged on a ring defined as a 2-BDG. A ring has
  been chosen in order to simulate an infinitely long chain of cells,
  avoiding the healing process to be triggered at the borders of the
  chain. When one cell of the ring is removed (or dies), the
  regulation results in the duplication of one of the cells at
  distance 2 from the removed cell. This is consistent with the
  observed biological behavior where division does not occur on the
  edges of a wound but one ``layer'' of cells farther.
\end{enumerate}

\section{Related and future works}

\begin{figure}[!b]
      \centering
      \includegraphics[width=0.72\linewidth]{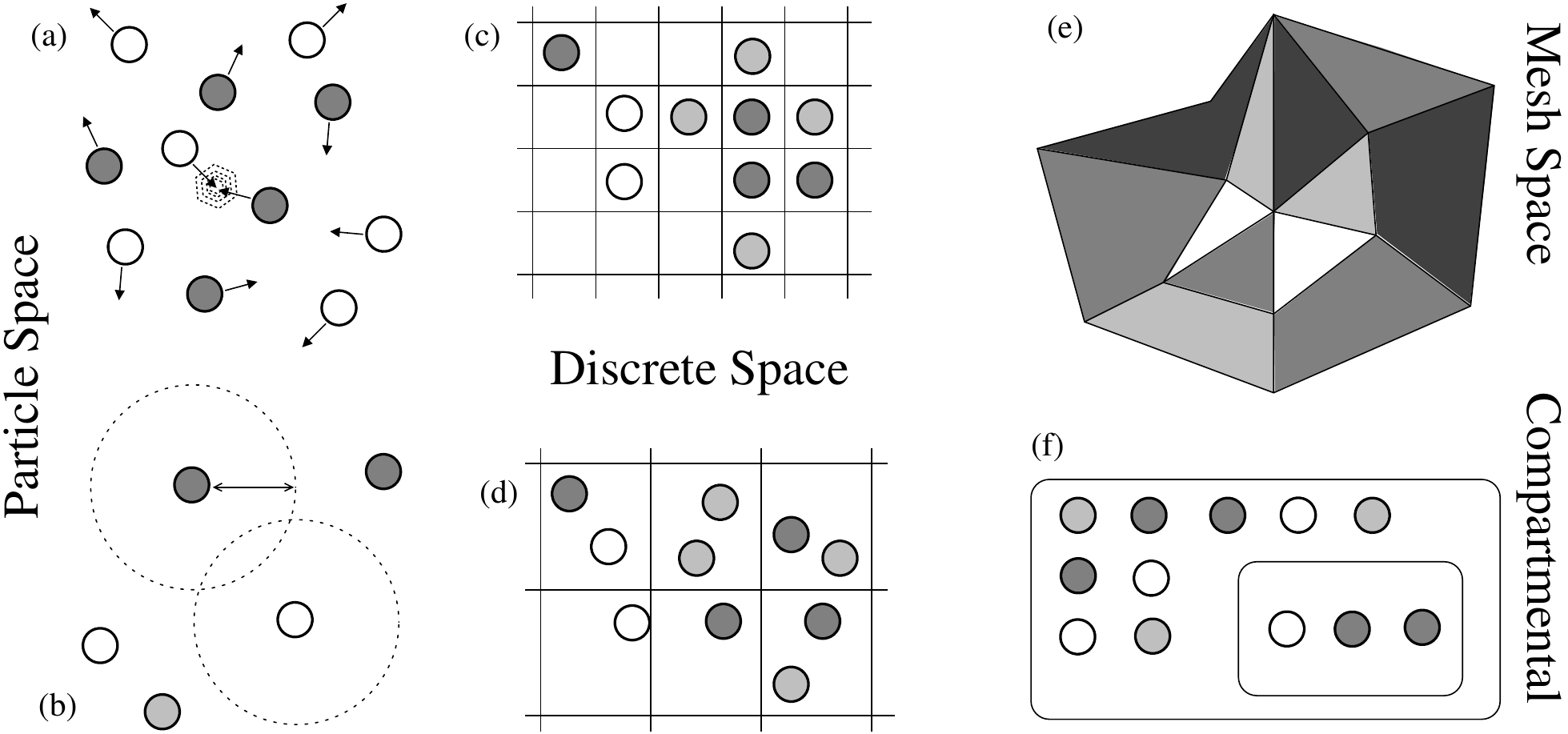}
      \caption{Various representation of space in systems biology
        according~\cite{Takahashi2005}.
        (a)~In particle space, molecules are represented as individual particles
        with positions in a continuum space. (b)~Such
        methods accommodates a discrete dynamics when particles jump in time
        and space by calculating the maximum distance that the particle can
        travel and interact in the time slot.
        Lattice discretizes regularly the space: in microscopic
        lattices~(c) at most one particle is allowed to occupy
        a lattice site while mesoscopic methods~(d) allows
        several.
        (e)~Mesh spaces are usually used to solve PDE (\eg, in
        reaction-diffusion systems).
        (f) Compartmental spaces focus on the molecular transfers
        between compartments.  
      }
      \label{fig:space}
\end{figure}

Integrative modelling of biological processes (\emph{e.g.} in system
biology) relates different models that operate on different levels of
abstraction and various spatial and time scales. For instance, in
developmental biology, the spatial organization of cells is especially
crucial and the description of the morphogenetic processes at a
cellular level~\cite{Reuille2006} implies the integration of molecular
mechanisms such as cell-cell signaling, mechanical stresses and
genetic regulation on a complex dynamic geometry.
Various formalism have been proposed to face these needs relying on
various explicit representations of space~\cite{Takahashi2005}:
molecular dynamics, spatial Gillespie~\cite{Stundzia1996}, partial
differential equations (PDE), lattice based methods (\eg, cellular
automata), etc. See Fig~\ref{fig:space} for a brief taxonomy of space
representation in biological modelling.

In this context, we want to develop a modelling framework suitable to
represent and study the regulations in multi-cellular systems, taking
into account the spatial relationships between the cells as well as
the spatial transformations resulting from cells divisions,
migrations, or apoptosis.
Discrete algebraic formalisms like P systems, process algebra or Petri
net are very relevant because the automated tools that can be used to
help both the modelling and the systematic analysis of the system
behavior.
Such formalisms take into account spatial relationships.  For
instance, classical membrane systems rely on membranes inclusion to
abstract the spatial organization of cellular processes. The
limitations of this structure has been recognized~\cite{giavitto02}
and leads to the development of several variants: tissue P
systems~\cite{vide2003tissue}, population P
systems~\cite{Bernardini:2004}, etc.
Some process algebra (\eg, used to study mobility or variants of
$\pi$-calculus used for biological modelling) rely also on a notion of
localization but often the spatial relationships are not explicitly
exposed (the algebra of locations is coded into identifiers) or too
limited (nesting structures).

The framework presented in this paper is based on the well-known
formalism of logical regulatory networks and extends it with spatial
information and a definition of the spatial transformations. A Petri
net semantics of the resulting formalism has been proposed in order to
allow for the simulation and the analysis of the modelled systems
(\eg, through model-checking, invariant extraction, etc.). 
Our solution is particularly flexible and general thanks to a
\emph{decoupling} of various concerns: in particular the regulation
aspects are grasped by regulatory networks while the spatial aspects
are handled through topological collections which unify several
spatial representations. These different aspects are addressed using
their own adequate tools. They are then related by means of a general
interface allowing a smooth bi-directional interaction while
preserving the decoupling.

The current presentation is limited to systems in which only one kind
of cell can exist. Our prototype implementation does not have such a
limitation since this is not a technical limitation but a deliberate
presentation decision to in order to focus on the intuition about our
approach. Future works will provide a generalized formal definition of
the framework presented in the paper. We also intend to run several
case studies in order to assess the relevance of our proposal: in
particular we shall consider models of \emph{Drosophilia} embryo as
in~\cite{ckp10} and extend them with cells divisions. Another aspect
will be addressed in the future: allowing that cells division
immediately results in the possibility of infinite state space
systems. This can be tackled by limiting the number of living cells at
a given time (\ie, use a finite set of module identifiers); another
possibility is to resort to abstraction, for instance by applying to
module identifiers a state space reduction technique originally
designed for systems with dynamic process
creation~\cite{ISCIS09,SACS10}. Finally, we intend to study an
alternative approach for spatial information in order to address
systems such as blood-cells populations. The idea is to implement the
spatial relation as a purely stochastic relation reflecting the idea
that such cells are in constant movement but may meet occasionally.

\smallskip

\noindent%
\emph{Acknowledgements.} The authors thank F. Delaplace at the
University of \'Evry for fruitful discussions and the anonymous
reviewers for their valuable comments.  This work is partially
supported by the ANR research projects AutoChem and Calamar.


\end{document}